\documentclass[12pt,a4paper]{article}
\usepackage[margin=2cm]{geometry}
\usepackage{comment}
\usepackage{amsmath,amssymb,extarrows,graphicx,subfigure,setspace}
\usepackage{cite}
\usepackage{slashed}
\usepackage{color}
\usepackage{braket}
\usepackage{float}
\makeatother
\usepackage{pdfpages}
\usepackage{blindtext}
\usepackage{amsfonts}
\usepackage{tensor}
\usepackage{graphicx}
\usepackage{pict2e}
\usepackage{amssymb, amsmath,environ}
\usepackage[
colorlinks=true,
linkcolor=blue,
urlcolor=blue,
filecolor=blue,
citecolor=red,
pdfstartview=FitV,
pdftitle={},
pdfauthor={},
pdfsubject={},
pdfkeywords={},
pdfpagemode=None,
bookmarksopen=true
]{hyperref}

\usepackage{epsfig}

\usepackage{hyperref}

\newcommand{\be}{\begin{equation}}
	\newcommand{\bea}{\begin{eqnarray}}
		\newcommand{\eea}{\end{eqnarray}}
	\newcommand{\ba}{\begin{array}}
		\newcommand{\ea}{\end{array}}
	\newcommand{\ee}{\end{equation}}
\newcommand{\bes}{\begin{equation*}}
	\newcommand{\beas}{\begin{eqnarray*}}
		\newcommand{\eeas}{\end{eqnarray*}}
	\newcommand{\bas}{\begin{array*}}
		\newcommand{\eas}{\end{array*}}
	\newcommand{\ees}{\end{equation*}}

\numberwithin{equation}{section}
\begin{document}
	\onehalfspacing
	\noindent
	
		\begin{titlepage}		\vspace{10mm}		\vspace*{20mm}
			\begin{center}			{\Large {\bf  Information Dynamics in Quantum Harmonic Systems: Insights from Toy Models}\\			}			\vspace*{15mm}			\vspace*{1mm}			{ Reza Pirmoradian $^{a,b}$ and M Reza Tanhayi $^{a,c}$}		
				
					\vspace*{.3cm}	{\it 				${}^a $ School of Quantum Physics and Matter,\\
						Institute for Research in Fundamental Sciences (IPM), P.O.Box 19395-5531,	Tehran, Iran		
						
						${}^b $ Ershad Damavand, Institute of Higher Education (EDI)\\ 				P.O. Box 14168-34311, Tehran, Iran}\\ 	{\it 				${}^c $ 	Department of Physics, Central Tehran Branch, Islamic Azad University  (IAU)\\ P.O. Box 14676-86831,				Tehran, Iran}\\
				{\it 			}	
						
				\vspace*{0.5cm}			{E-mails: {\tt rezapirmoradian@ipm.ir, mtanhayi@ipm.ir}} 			\vspace*{1cm}
\end{center}	\date{\today}		\begin{abstract} 

This study investigates the dynamics of quantum information and computational resources using a tractable model of coupled harmonic oscillators. We precisely characterize the interplay between mutual information, synchronization, and circuit complexity, demonstrating that they serve as complementary yet distinct measures of quantum correlations. Our analysis reveals how coupling strength, detuning, and external magnetic fields modulate these quantities, with synchronization and mutual information exhibiting marked divergence in nonlinear regimes. By employing exact Gaussian methods, we compute the circuit depth required to prepare target states and connect increased fidelity to more regular dynamical behavior. Furthermore, we analyze single-ion transport in a harmonic trap, comparing sudden and adiabatic protocols. We introduce a nonadiabaticity metric to quantify the fidelity-complexity trade-off, showing that smooth control sequences significantly minimize operational errors by suppressing excitations. These results provide a refined understanding of quantum correlations and offer concrete principles for optimizing control strategies in quantum technologies.

			
			
			\end{abstract}	\end{titlepage}

				\tableofcontents


\section{introduction}	
Harmonic oscillators constitute a cornerstone of quantum mechanics, providing an analytically tractable framework for probing complex phenomena such as entanglement dynamics and quantum control. Their Gaussian structure enables exact solutions, making them a vital testbed for advancing and refining theoretical approaches. This utility is particularly evident in coupled-oscillator models, which serve as direct analogues for leading quantum technological architectures. A canonical example is the ion trap, proposed by Cirac and Zoller \cite{c1}, where inter-ion interactions are naturally described by coupled oscillators \cite{c2}. Within such systems, synchronization emerges as a critical mechanism for sustaining coherent control, suppressing decoherence, and enhancing the fidelity of entangling quantum gates \cite{Bruzewicz:2019}.

A principal challenge in these systems lies in understanding the intricate interplay between quantum correlations, operational fidelity, and the resulting computational complexity. Within the Hamiltonian framework, key metrics offer crucial insights into this balance. Entanglement entropy, for instance, quantifies quantum correlations between subsystems, providing insight into their shared information. This measure is foundational to quantum information science, with direct applications in areas such as error correction and cryptography. A second critical metric is computational complexity, often quantified by circuit depth, the number of sequential quantum gate layers required to implement a computation\footnote{Circuit depth, used here as a proxy for quantum complexity, is interpreted within the Nielsen geometric framework unless otherwise stated. (for more theoretical details see \cite{Susskind:2014moa, Susskind:2014rva, Brown:2015lvg, Chapman:2021jbh, Headrick:2019eth, Alishahiha:2019cib, Belin:2021bga, RezaTanhayi:2018cyv, Alishahiha:2018lfv, Alishahiha:2014jxa, Borvayeh:2020yip, Tanhayi:2015cax}). 
}. While greater circuit depth can enhance algorithmic power, it also increases susceptibility to decoherence and operational errors.

A fundamental, yet incompletely understood, relationship exists between these metrics. Generally, higher entanglement entropy correlates with greater computational complexity, as maintaining robust quantum correlations demands more intricate and deeper circuits. This interplay is central to the Nielsen geometric framework \cite{Susskind:2014moa, Brown:2015lvg}, which provides a rigorous measure for the minimal resources, or complexity, required to implement a quantum operation. Initial investigations have begun to chart this relationship in specific contexts; for instance, the work of Ameri et al. \cite{amer} established a connection between quantum synchronization and mutual information. Nevertheless, a comprehensive, quantitative mapping of how this interplay is governed by fundamental control parameters, such as coupling strength and external fields, remains an open area of inquiry. The consequent implications for the scaling of complexity are yet to be firmly established.

In this work, we test and extend the proposed link between synchronization and information theory using a toy model of two coupled harmonic oscillators. We demonstrate that the relationship is more intricate than previously recognized: strong coupling can increase mutual information while simultaneously degrading synchronization, and external magnetic fields suppress both quantities. To quantify the operational cost of these quantum correlations, we employ the Nielsen geometric framework \cite{Chapman:2021jbh, Headrick:2019eth}, interpreting circuit depth as a measure of computational complexity.

We further explore these dynamics through a quantum quench protocol, tracking the temporal evolution of synchronization and mutual information. Finally, to connect our results to a broader quantum informational context and foundational complexity measures \cite{Alishahiha:2019cib, Belin:2021bga}, we examine fidelity and complexity within the thermofield double state (TFD) formalism, leveraging the Gaussian nature of our states to compute complexity via the covariance matrix method.

This paper is structured as follows: In Section 2, we develop the theoretical framework for a two-body system of coupled harmonic oscillators. Section 3 investigates the core interplay between coupling strength and external fields, presenting our central results on synchronization, mutual information, and circuit complexity. Section 4 analyzes a complementary one-body model of ion transport to elucidate fundamental principles of quantum control and fidelity. We conclude with a synthesis of our findings and their implications in Section 5.


\section{A Model of a Two-Body System: Coupled Harmonic Oscillators}

The dynamics of ions confined in trapping potentials are often well-approximated by harmonic oscillators, particularly near the potential minimum. While a single-particle model offers initial insights, realistic quantum technologies like ion traps involve multiple ions whose motions are coupled via Coulomb interactions. These interactions give rise to collective motional modes, necessitating a coupled-oscillator description. This section develops a two-body coupled harmonic oscillator model to analyze these effects, incorporating an external magnetic field to study its influence on synchronization and information dynamics.

\subsection{Hamiltonian Formulation}
We begin with the standard Hamiltonian for two coupled harmonic oscillators in a quantum framework:
\begin{equation}\label{ham}
	H = \omega_1 \hat{a}^\dagger \hat{a} + \omega_2 \hat{b}^\dagger \hat{b} + g' (\hat{a} + \hat{a}^\dagger)(\hat{b} + \hat{b}^\dagger),
\end{equation}
where \(\omega_1\) and \(\omega_2\) are the natural frequencies, \(g'\) is the coupling strength, and \(\hat{a}, \hat{a}^\dagger\), \(\hat{b}, \hat{b}^\dagger\) are the annihilation and creation operators for the two modes, respectively. The interaction term \(g' (\hat{a} + \hat{a}^{\dagger}) (\hat{b} + \hat{b}^{\dagger})\) represents a linear coupling that facilitates energy exchange between the oscillators.
To investigate the impact of an external field, we introduce a magnetic field using the symmetric gauge \(\vec{A} = \frac{B}{2} (x_2 - x_1)\). This gauge choice preserves the analytical tractability of the model while capturing essential physics. Transforming to position and momentum operators (\(x_j, p_j\)) and setting \(m=1\), the Hamiltonian becomes:
\begin{equation}
	H = \frac{1}{2} \sum_{j=1}^{2} \left(p_{j}^{2} + (\omega_{j}^{2} + \omega_{c}^{2}) x_{j}^{2} \right) - g x_1 x_2 + \omega_c \left( p_1 x_2 - p_2 x_1 \right),
	\label{eq:hamiltonian_magnetic}
\end{equation}
where \(\omega_c = eB/(2c)\) is the cyclotron frequency and \(g = \frac{1}{2} g' \sqrt{\omega_1 \omega_2}\) is the redefined coupling parameter. The magnetic field introduces two key modifications: a frequency shift \(\omega_c^2\) and a momentum-position coupling term \(\omega_c (p_1 x_2 - p_2 x_1)\), which is crucial for synchronization dynamics.

To simplify the Hamiltonian in Eq.~\eqref{eq:hamiltonian_magnetic}, we apply a time-dependent canonical transformation to new coordinates \((X_j, P_j)\):
\begin{equation}
	H' = H - \dot{\phi}(t) (P_{1}X_{2} - P_{2}X_{1}).
	\label{eq:transformed_hamiltonian}
\end{equation}
By selecting \(\dot{\phi}(t) = \omega_c\), the generating function of this transformation eliminates the cross-term \( \omega_c (p_1 x_2 - p_2 x_1) \) from the original Hamiltonian. The solution \(\phi(t) = \omega_c t + \theta\) follows directly, with \(\theta\) as an integration constant.
In the rotated frame, the Hamiltonian simplifies to:
\begin{equation}
	H' = \frac{1}{2}\sum_{j=1}^{2}\left(P_{j}^{2} + \Omega_{j}^{2}(t) X_{j}^{2}\right) + \Omega^2_{12}(t) X_{1}X_{2},
	\label{eq:hamiltonian_rotated}
\end{equation}
with time-dependent frequencies given by:
\begin{align}
	\Omega_{1}^{2}(t) &= \omega_{1}^{2}\cos^{2}\phi(t) + \omega_{2}^{2}\sin^{2}\phi(t) + \omega_c^2 + g\sin 2\phi(t), \nonumber\\
	\Omega_{2}^{2}(t) &= \omega_{1}^{2}\sin^{2}\phi(t) + \omega_{2}^{2}\cos^{2}\phi(t) + \omega_c^2 - g\sin 2\phi(t), \nonumber\\
	\Omega^2_{12}(t) &= \frac{\omega_{1}^{2} - \omega_{2}^{2}}{2}\sin 2\phi(t) - g\cos 2\phi(t).
	\label{eq:frequencies}
\end{align}
The system decouples instantaneously when the interaction term vanishes, \(\Omega_{12}(t) = 0\). This occurs at a specific mixing angle determined by:
\begin{equation}
	\tan 2\phi(t) = \frac{2g}{\omega_1^2 - \omega_2^2}.
	\label{eq:decoupling_condition}
\end{equation}
This condition defines the precise relationship between the magnetic field (through \(\phi(t)\)), oscillator frequencies, and coupling strength required for momentary decoupling. For constant parameters, complete permanent decoupling is generally not possible due to the explicit time dependence in \(\phi(t)\).

\subsection{Time-Dependent Wavefunction}
The ground state wavefunction for this time-dependent system maintains a Gaussian form, preserved under the action of scaling and entangling operations \cite{jalal0, jalal2, Hab-arrih:2019xpe}. The general solution to the time-dependent Schrödinger equation is:
\begin{equation}
	\Psi(x_{1}, x_{2}; t) = \mathcal{N} \exp\left[-\frac{1}{2}\left(A_{1}x_{1}^{2} + A_{2}x_{2}^{2} - A_{12}x_{1}x_{2}\right)\right],
	\label{eq:wavefunction}
\end{equation}
where the normalization factor is:
\begin{equation}
	\mathcal{N} = \left(\prod_{j=1}^{2} \frac{\Omega_{j}(0)}{\pi^{2} h_{j}^{2}(t)}\right)^{\frac{1}{4}} \exp\left[-\frac{i}{2}\left(\Omega_{1}(0)\int_{0}^{t}\frac{dt'}{h_{1}^{2}(t')} + \Omega_{2}(0)\int_{0}^{t}\frac{dt'}{h_{2}^{2}(t')}\right)\right].
\end{equation}
The functions \(h_j(t)\) satisfy the Ermakov equation:
\begin{equation}
	\ddot{h}_{j} + \Omega_{j}^{2}(t) h_{j} = \frac{\Omega_{j}^{2}(0)}{h_{j}^{3}},
\end{equation}
with initial conditions \(h_j(0) = 1\) and \(\dot{h}_j(0) = 0\). The time-dependent coefficients in the wavefunction are:
\begin{align}
	A_1 &= \left(\frac{\Omega_1(0)}{h_1} - i\frac{\dot{h}_1}{h_1}\right)\cos^2\phi(t) + \left(\frac{\Omega_2(0)}{h_2} - i\frac{\dot{h}_2}{h_2}\right)\sin^2\phi(t), \nonumber\\
	A_2 &= \left(\frac{\Omega_2(0)}{h_2} - i\frac{\dot{h}_2}{h_2}\right)\cos^2\phi(t) + \left(\frac{\Omega_1(0)}{h_1} - i\frac{\dot{h}_1}{h_1}\right)\sin^2\phi(t), \nonumber\\
	A_{12} &= \left[\frac{\Omega_1(0)}{h_1} - \frac{\Omega_2(0)}{h_2} - i\left(\frac{\dot{h}_1}{h_1} - \frac{\dot{h}_2}{h_2}\right)\right]\sin\phi(t)\cos\phi(t).
	\label{eq:coefficients}
\end{align}
This Gaussian state formalism provides the foundation for our subsequent analysis of quantum metrics, including synchronization, mutual information, and circuit complexity, which will be explored in the following sections.

\subsection{Quench and Steady-State Approximation}
The quench model involves the sudden change of system parameters—such as interaction strength or local frequencies to probe non-equilibrium dynamics, a key aspect in understanding quantum phase transitions, entanglement growth, and thermalization. Such dynamics are especially relevant in regimes where classical simulations become intractable \cite{qu}. In this work, we consider a realistic quench protocol inspired by experimentally feasible settings. Specifically, at $t = 0$, the local mode frequencies $\omega_1$, $\omega_2$, and the coupling strength $g$ are instantaneously changed from initial constant values to new final constants:
\be
\omega_j = \left\{
\begin{array}{ll}
	\omega_{ij} & t = 0 \\
	\omega_{fj} & t > 0 \\
\end{array}
\right., \quad
g = \left\{
\begin{array}{ll}
	0 & t = 0 \\
	g & t > 0 \\
\end{array}
\right.
\ee
where $j = 1, 2$. This form of global quench is not only analytically tractable but also experimentally realizable. In particular, sudden changes in trapping frequencies or coupling rates are routinely implemented in platforms such as trapped ion systems and superconducting circuit QED architectures. In these settings, external control parameters, like laser intensities or flux-tunable couplers, can be rapidly modulated with high precision, making the adopted quench model both theoretically meaningful and experimentally relevant. In this scenario, the solutions of the {Ermakov} equations  now take the forms
\begin{eqnarray}
	h_{1}^{2}(t) &=&\frac{\Omega_{f1}^{2}-\Omega_{i1}^{2}}{2\Omega_{f1}^{2}}\cos\left(2\Omega_{f1}\ t\right)+\frac{\Omega_{f1}^{2}+\Omega_{i1}^{2}}{2\Omega^2_{f1}}\nonumber\\ 
	h_{2}^{2}(t) &=&\frac{\Omega_{f2}^{2}-\Omega_{i2}^{2}}{2\Omega_{f2}^{2}}\cos\left(2\Omega_{f2}\ t\right)+\frac{\Omega_{f2}^{2}+\Omega_{i2}^{2}}{2\Omega_{f2}^{2}}.\label{h}
\end{eqnarray} 
The problem is simplified to the ground state of two coupled harmonic oscillators.

To make connection with the next section, it is important to introduce the steady-state approximation, a key idea that simplifies how we analyze system behavior. This method works under certain conditions: the system must change slowly compared to its natural frequencies $\Omega_{f1}$ and $\Omega_{f2}$ to avoid sudden transitions, a strong external field $\omega_c$ must help maintain the mixing angle $\theta$, and the functions $h_j(t)$ (Eq. \ref{h}) should settle into steady values, reducing temporary fluctuations. When these conditions are met, the simplified depth formula and its scaling rules remain valid. If not, time-dependent factors like $\dot{h}_j/h_j$ in $A_j(t)$ create unpredictable oscillations, making both theoretical studies and practical applications more challenging. The steady-state model offers valuable insights into the complexity of control systems, correctly predicting long-term depth behavior and identifying universal scaling rules, such as the logarithmic relationship with $\omega_c$. Under this approximation, the wave function is given by \eqref{eq:wavefunction}, where
\begin{eqnarray}\label{st}
	&& A_{1}=\Omega_{1}\cos^{2}\theta+\Omega_{2}\sin^{2}\theta
	,\hspace{.5cm} A_{2}=\Omega_{2}\cos^{2}\theta+\Omega_{1}\sin^{2}%
	\theta\nonumber \\
	&& A_{12}=\Big(\Omega_{1}-\Omega_{2}\Big)\sin\theta\cos\theta,\nonumber\\
	&&
	\Omega_{1}   =\sqrt{\omega_{1}^{2}\cos^{2} \theta
		+\omega_{2}^{2}\sin^{2} \theta+\omega_c^2 +g\sin 2\theta} ,\nonumber\\
	&&
	\Omega_{2}  =\sqrt{\omega_{1}^{2}\sin^{2}\theta  +\omega_{2}^{2}\cos^{2}\theta+\omega_c^2  -g\sin 2\theta }
\end{eqnarray}
the normalization constant is given by:  
$$  
\mathcal{N} = \left( \frac{\Omega_{1} \Omega_{2}}{\pi^{2}} \right)^{\frac{1}{4}},  
$$  
noting that when \(\omega_1 = \omega_2\), the angle \(\theta\) simplifies to \(\frac{\pi}{4}\).

\section{Quantum Correlations and Computational Complexity}

This section analyzes the interplay between quantum correlations and computational resources in our two-body system. We demonstrate that synchronization and mutual information serve as complementary correlation measures exhibiting distinct dynamical signatures, while circuit depth quantifies the computational overhead required to maintain these quantum correlations.

\subsection{The Synchronization-Information Contrast}

Quantum synchronization is the phenomenon where the expectation values of quantum observables, such as position and momentum quadratures in continuous, variable systems, exhibit stable, correlated temporal evolution. To quantify this phenomenon the following measure is often used that evaluates relative phase-space localization and coherence \cite{bo,Galve-2017,mar}:

\begin{equation}
	\mathcal{S}_c = \frac{1}{\langle (\hat{p}_1 - \hat{p}_2)^2 \rangle + \langle (\hat{x}_1 - \hat{x}_2)^2 \rangle},
	\label{eq:sync}
\end{equation}
where enhanced synchronization corresponds to minimized position-momentum differences between subsystems.

Complementarily, mutual information quantifies the total correlations between two systems by measuring the reduction in uncertainty about one system given knowledge of the other, defined as:
\begin{equation}
	I_{AB} = S(\rho_A) + S(\rho_B) - S(\rho_{AB}),
	\label{eq:mutual_info}
\end{equation}
where \(S(\rho) = -\text{Tr}(\rho \log \rho)\) denotes the von Neumann entropy, with \(\rho_{AB}\) the density matrix of the composite system, and \(\rho_A = \operatorname{Tr}_B(\rho_{AB})\), \(\rho_B = \operatorname{Tr}_A(\rho_{AB})\) the reduced density matrices of subsystems \(A\) and \(B\) respectively.

Synchronization and mutual information provide complementary, yet distinct, characterizations of quantum correlations: the former quantifies phase-space alignment through the second-order moments of quadrature operators, making it a natural measure for Gaussian states, while the latter, derived from the von Neumann entropy, captures the total statistical correlations, both classical and quantum, between subsystems. In Gaussian systems, synchronization and mutual information are directly related, as both are determined solely by the covariance matrix. However, this equivalence breaks down under the influence of nonlinear interactions, such as those introduced by a Kerr-type nonlinearity in the Hamiltonian \eqref{ham}, which generate non-Gaussian dynamics. In such regimes, the growth of complex, higher-order correlations captured by mutual information can diverge from the simple phase-space coordination reflected by synchronization, revealing a fundamental split between informational and dynamical pictures of correlation.

\begin{figure}
	\includegraphics[width=8.5cm]{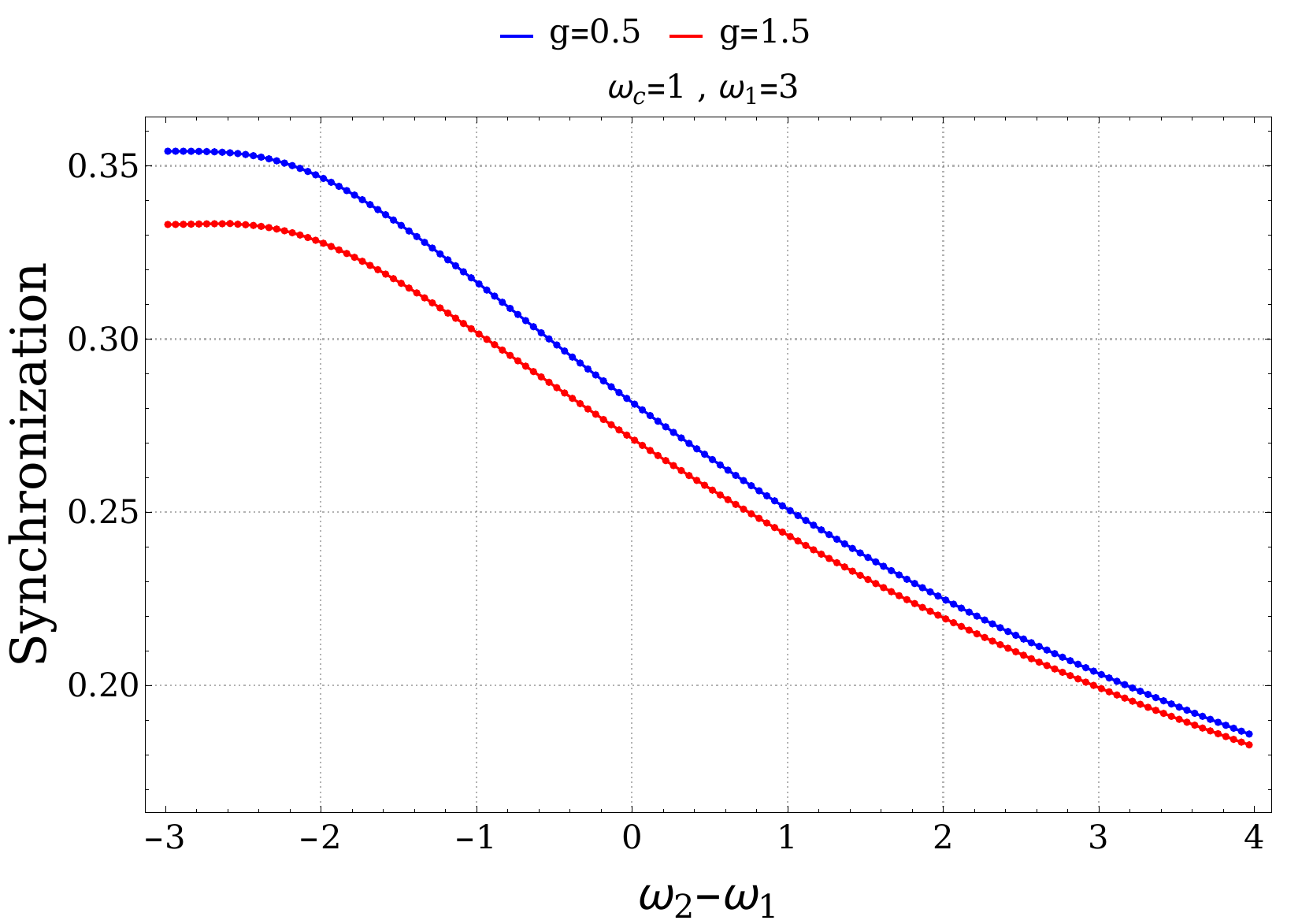} \includegraphics[width=8.5cm]{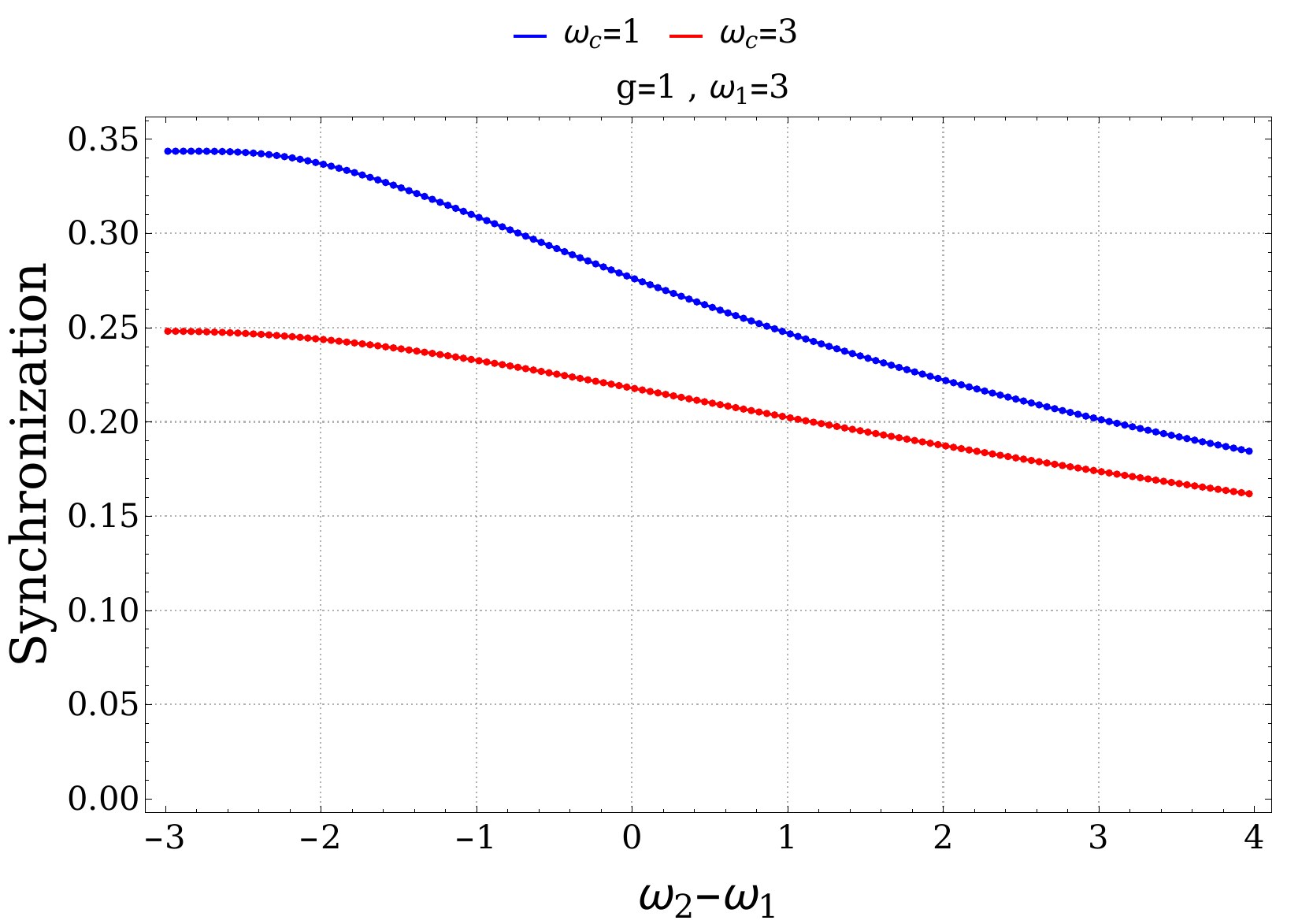} 
	\includegraphics[width=8.5cm]{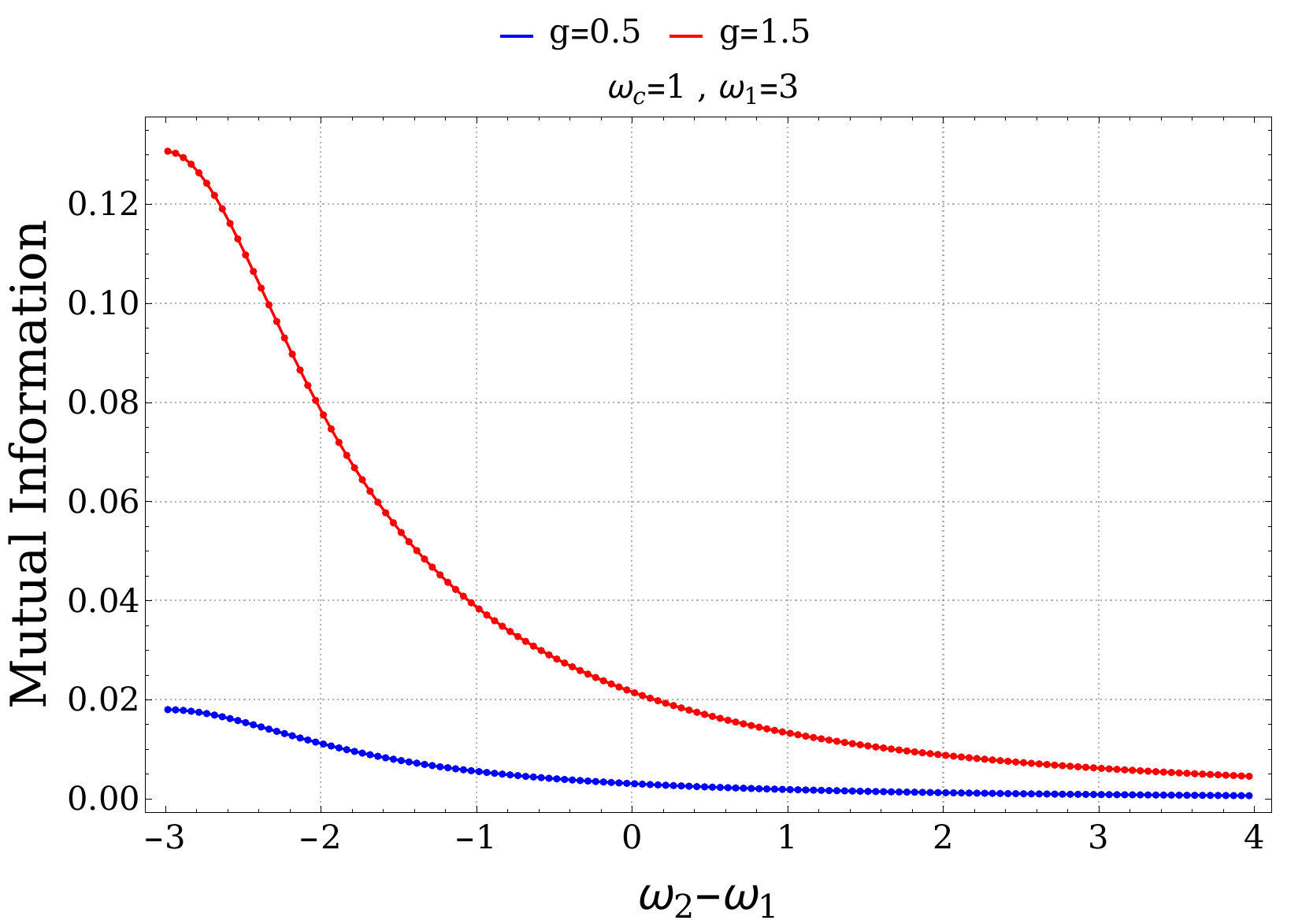} 	\includegraphics[width=8.5cm]{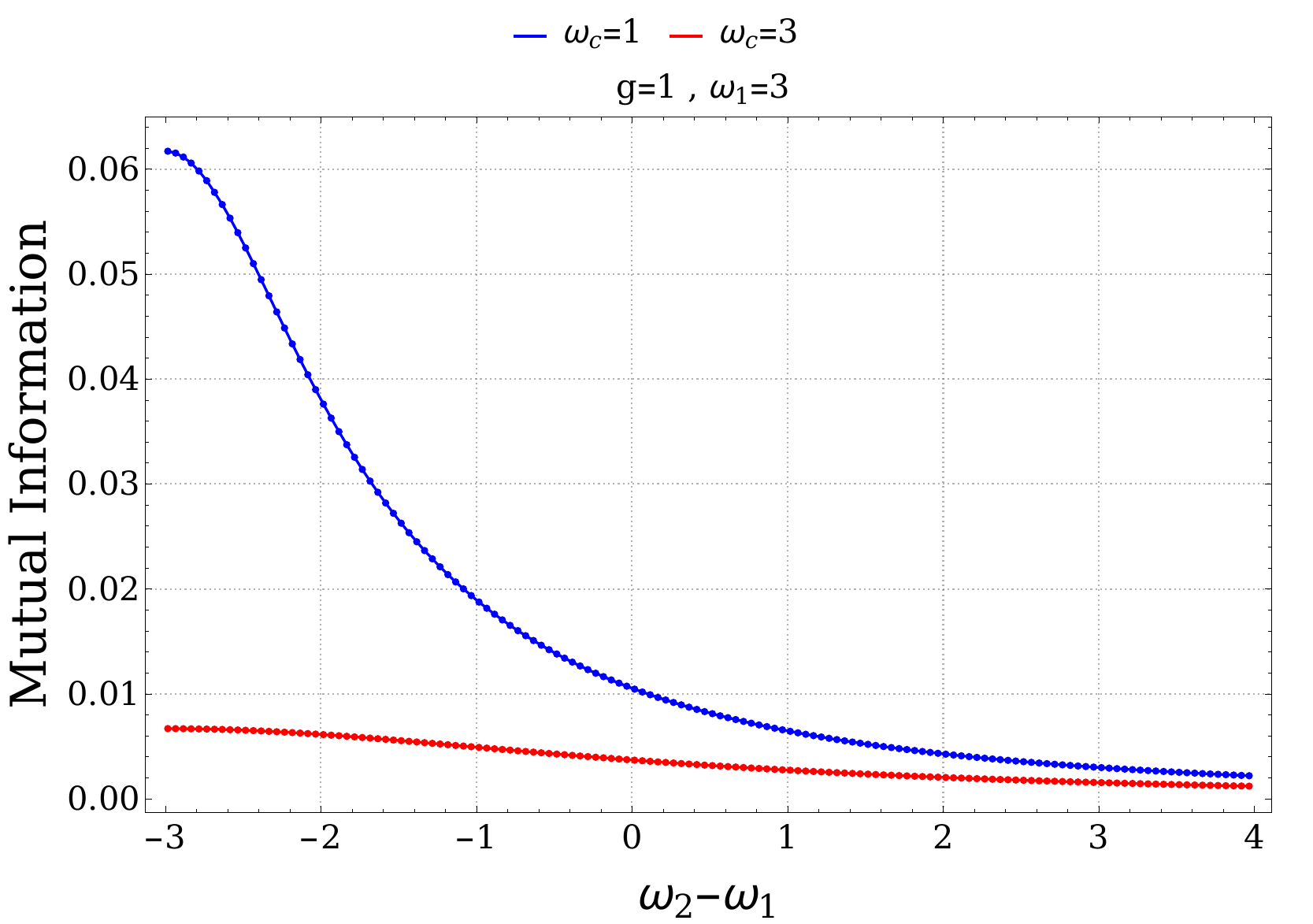}	
	\caption{ 
		The plots show the variation of synchronization (top) and mutual information (bottom) as a function of the frequency detuning $\omega_2 - \omega_1$ for different values of the coupling strength $g$ and cutoff frequency $\omega_c$. In the top panel, for $\omega_c = 1$, synchronization is shown for $g = 0.5$ (blue) and $g = 1.5$ (red). In the bottom panel, for $g = 1$, mutual information is plotted for $\omega_c = 1$ (blue) and $\omega_c = 3$ (red).
	}
	\label{Fig.3}
\end{figure}

Our study of coupled quantum oscillators provides a concrete manifestation of this fundamental split driven by the nonlinear coupling induced by the external magnetic field. We find that under specific nonlinear dynamics, controlled here by coupling strength, the behaviors of synchronization and mutual information diverge, cleaving the informational picture of correlation from the dynamical one.
\begin{itemize}
	\item Effect of coupling $g$: The left panel of Figure \ref{Fig.3} shows how the two measures react differently to stronger coupling. As the coupling gets stronger (from the blue to the red line), the total shared information (mutual information) increases and its curve becomes broader. However, the coordinated movement between the oscillators (synchronization) actually gets weaker and its peak becomes narrower.
		 
	\item Effect of magnetic field $\omega_c$: For a fixed coupling strength,  the right panel of Figure \ref{Fig.3} indicates that a stronger magnetic field (red curve) induces pronounced oscillations in the mutual information while simultaneously lowering the overall level of synchronization.
	
\end{itemize}

\begin{figure}
	\includegraphics[width=8.5cm]{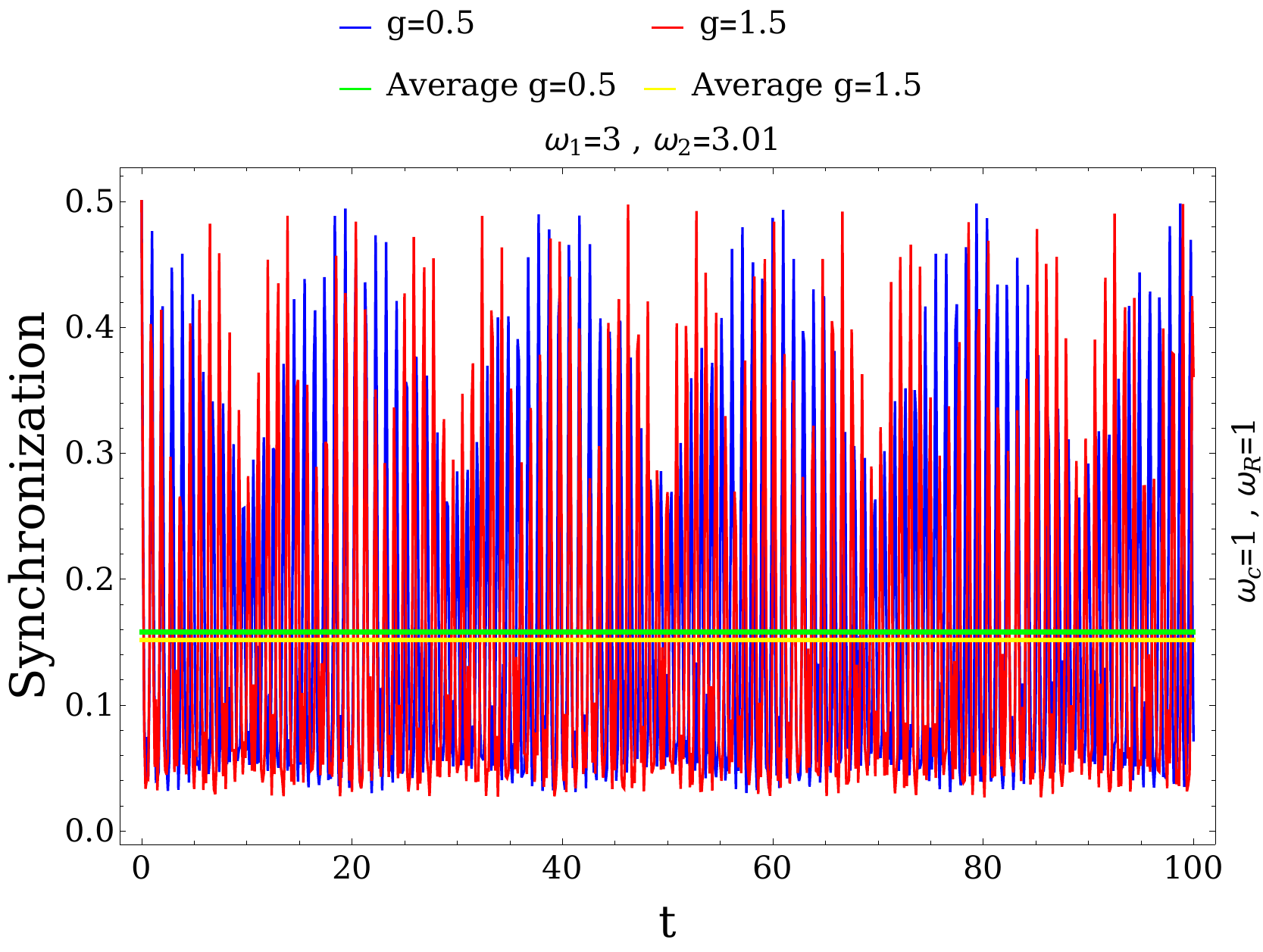}
	\includegraphics[width=8.5cm]{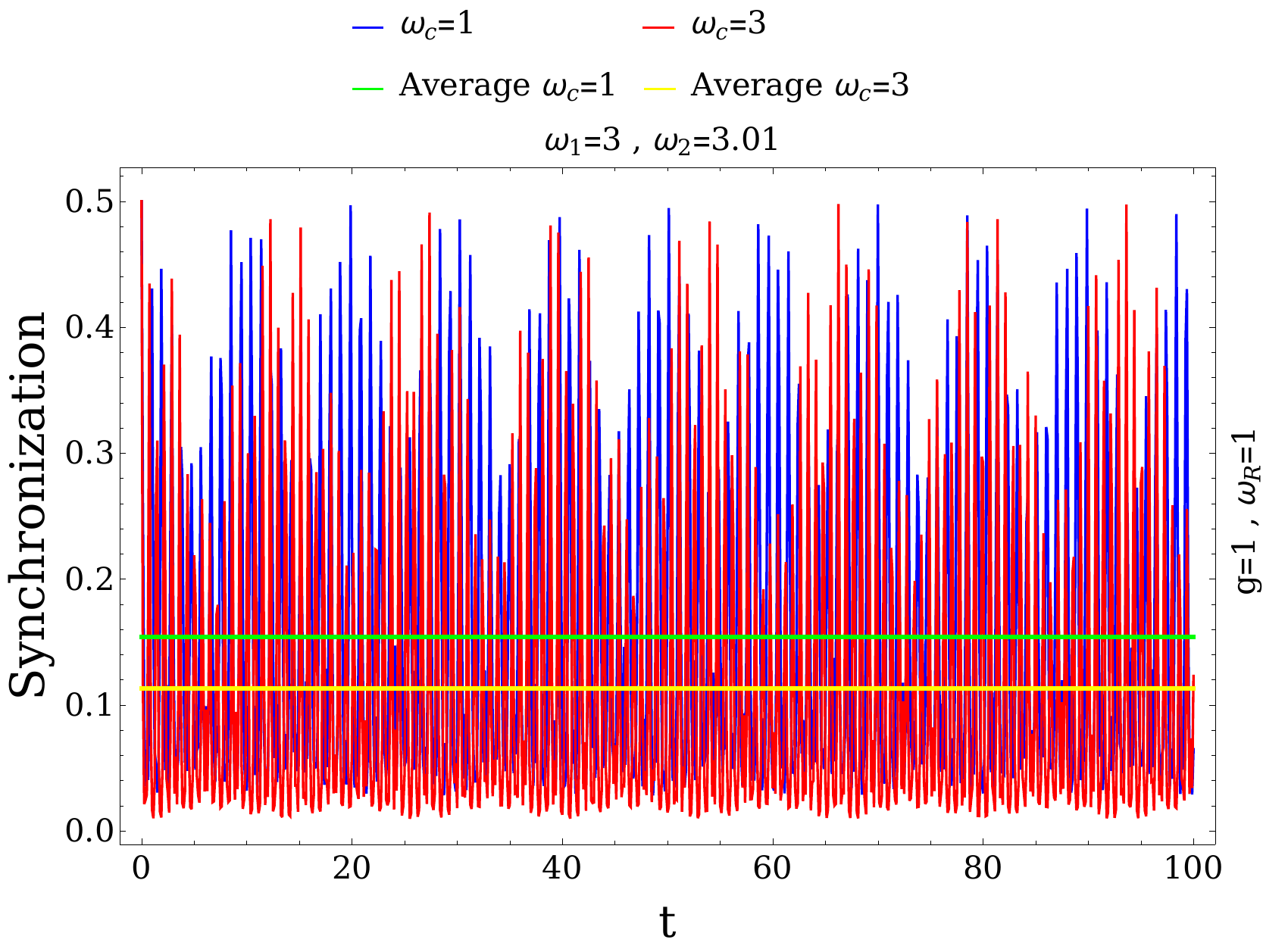} 
	\includegraphics[width=8.5
	cm]{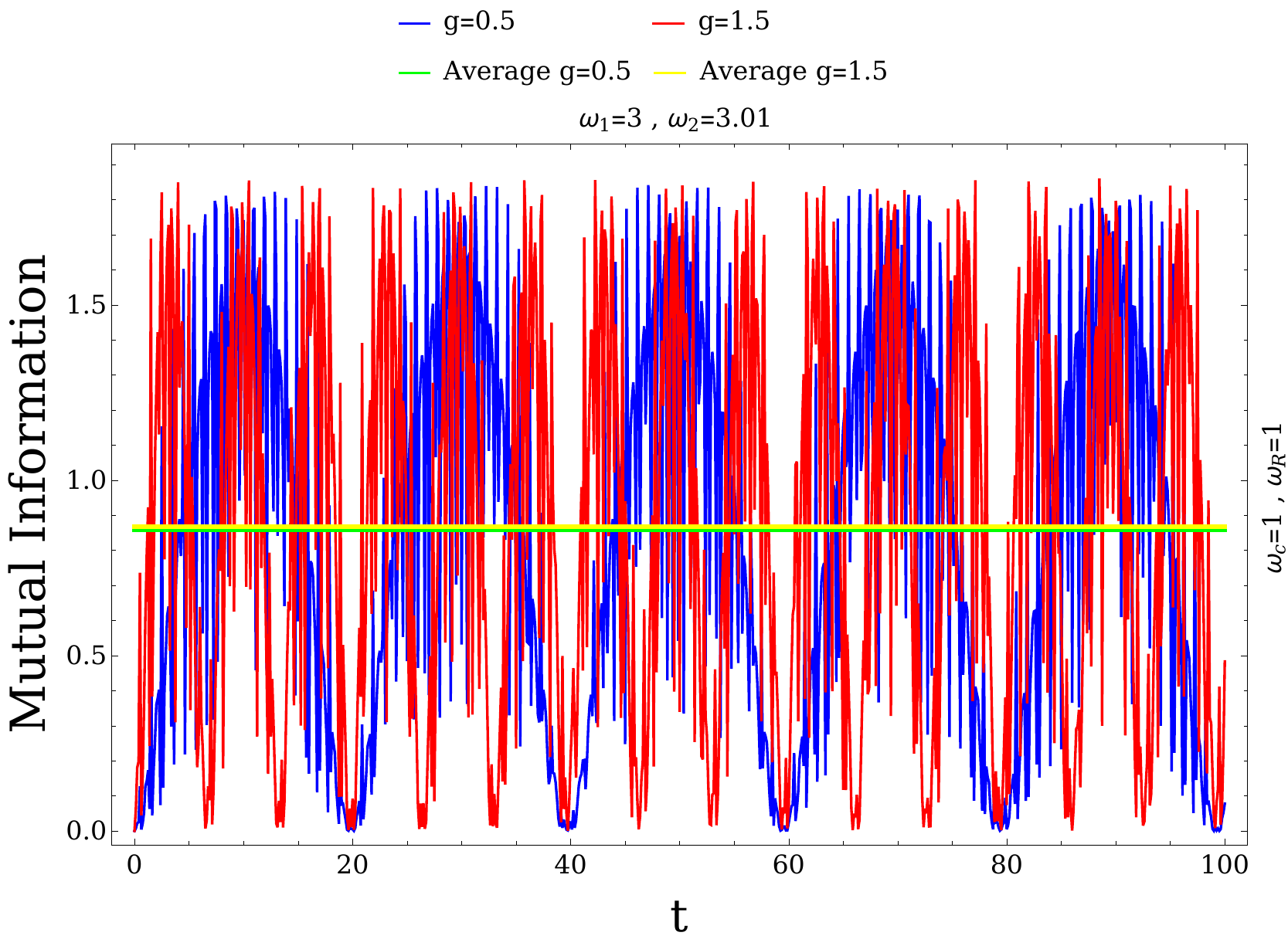}
	\includegraphics[width=8.5cm]{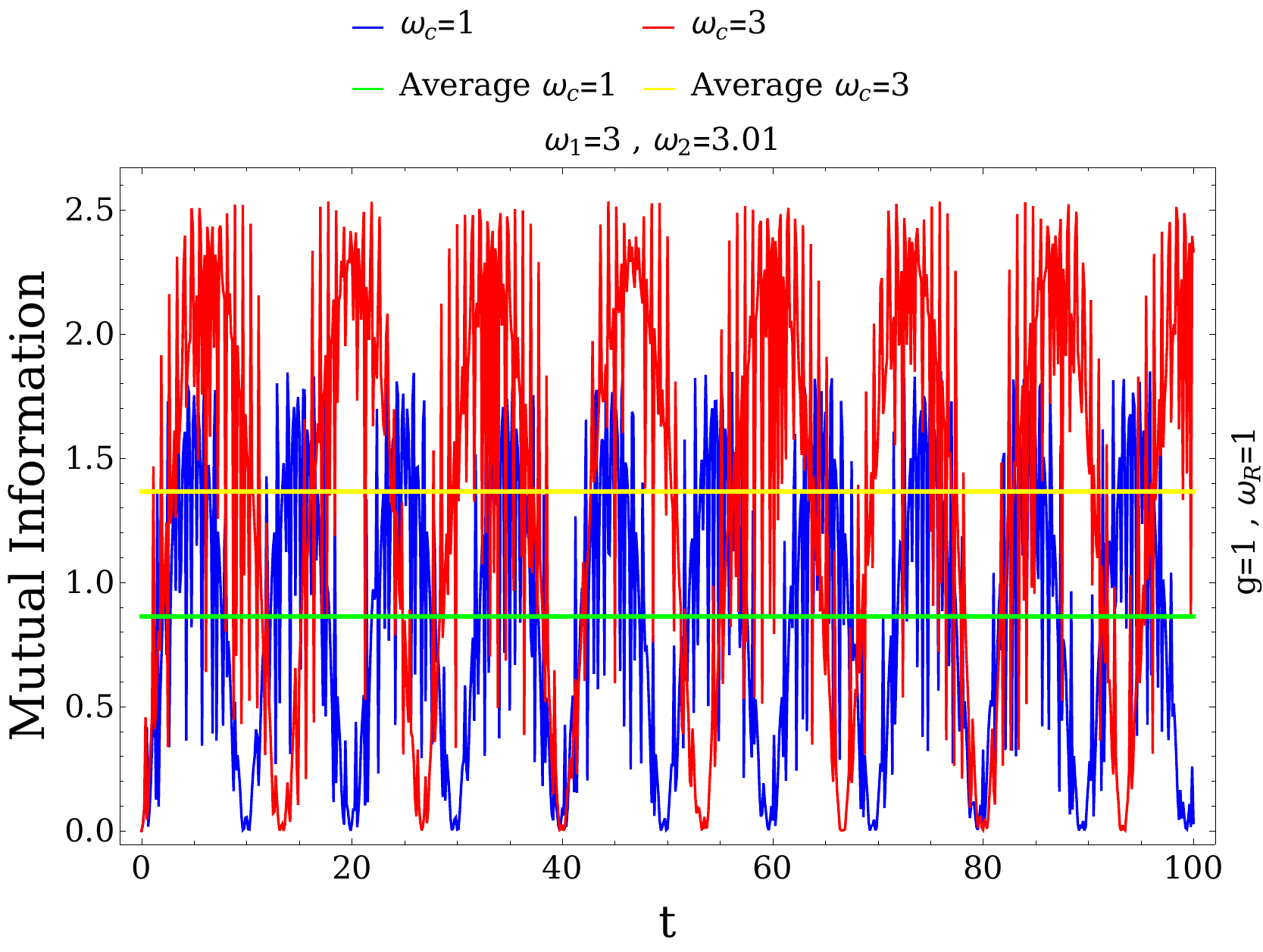}
	\caption{ 
		Time evolution of synchronization (top) and mutual information (bottom) for the quench model under various conditions. In the top panel, for $g = 1$ (blue) and $g = 1.5$ (red), synchronization is shown with frequency $\omega_c$ varying between 1 (blue) and 3 (red). In the bottom panel, mutual information is depicted for $g = 1$ (blue) and $g = 1.5$ (red), with frequency $\omega_c$ set to 1 (blue) and 3 (red). The plots also highlight the average values of both synchronization and mutual information over time.
	}	
	\label{Fig.5}
\end{figure}

A time-resolved analysis  reveals a key distinction between the roles of coupling strength and external magnetic fields. While stronger coupling increases the oscillatory behavior of both synchronization and mutual information without altering their long-term averages, the magnetic field fundamentally shifts the equilibrium between these measures. As shown in Figure \ref{Fig.5}, different coupling strengths produce distinct dynamical traces but similar averaged values. In contrast, a stronger magnetic field not only modifies the oscillatory pattern but also actively decouples the averages, suppressing time-averaged synchronization while enhancing time-averaged mutual information. This demonstrates that the magnetic field acts as a direct control parameter that can break the conventional balance between dynamical coordination and shared information.

Figure \ref{Fig.3} and Figure \ref{Fig.5} provide complementary perspectives on the interplay between synchronization and mutual information. Figure \ref{Fig.3} establishes the steady-state dependence of both measures on coupling strength and magnetic fields, revealing that stronger coupling increases mutual information while reducing synchronization, a divergence that becomes more pronounced under stronger magnetic fields. Figure \ref{Fig.5} tracks the time evolution of these quantities following a quench, showing dynamically how this steady-state relationship unfolds. A key distinction emerges: while stronger coupling primarily enhances oscillatory behavior, the magnetic field fundamentally alters the long-term balance between the two measures, sustaining higher average mutual information while suppressing synchronization. Together, the figures move from a static, parameter-space characterization to a dynamic confirmation of their controlled separation, underscoring that synchronization and mutual information offer complementary insights into coherent dynamics and shared information, respectively. Their combined use is essential for a complete understanding of quantum correlations in systems influenced by external fields and nonlinearities.

Building on this characterization, our results contrast with certain prior studies, such as Ameri et al. \cite{amer}, who observed synchronized behavior between these measures in driven-dissipative systems and proposed mutual information as a proxy for synchronization. Our findings underscore that the relationship is system-dependent: while mutual information robustly quantifies total correlations, its correspondence with synchronization is not guaranteed and is instead shaped by specific interaction forms and environmental couplings.

\subsection{Circuit Depth as a Measure of Quantum Complexity}

Having established the behavior of quantum correlations, we now turn to their computational cost. Quantum circuit depth, defined as the minimum number of sequential quantum gate layers required to implement a specific unitary operation, serves as a fundamental metric for computational efficiency and resource estimation. Within the Nielsen geometric framework, it provides a quantitative measure of quantum complexity, reflecting the minimal computational cost of preparing a target state from a reference state. While deeper circuits enable more complex state manipulations, they also increase susceptibility to decoherence and operational errors, presenting a critical trade-off in quantum algorithm design.

We evaluate the circuit depth for the transformation from a reference state \(|\psi_R\rangle\) to a target state \(|\psi_T\rangle\) via unitary evolution:
\[
|\psi_T\rangle = U|\psi_R\rangle,
\]
where the unitary \(U\) is decomposed into a sequence of fundamental quantum gates:
\[
U = \mathcal{O}_n \mathcal{O}_{n-1} \ldots \mathcal{O}_2 \mathcal{O}_1.
\]
The reference state is chosen as the factorized Gaussian ground state of two decoupled oscillators:
\begin{equation}
	\psi_R(x_1,x_2) = \sqrt{\frac{m\omega_R}{\pi}} \exp\left[-\frac{m\omega_R}{2}(x_1^2 + x_2^2)\right],
\end{equation}
with reference frequency \(\omega_R\). The target state is the time-evolved Gaussian wavefunction given in Eq.~\eqref{eq:wavefunction}. The gate set consists of local scaling operators \(\mathcal{O}_{jj} = e^{i\epsilon(x_jp_j + p_jx_j)/2}\) and non-local entangling gates \(\mathcal{O}_{ij} = e^{i\epsilon x_ip_j}\) (\(i \neq j\)), which collectively preserve the Gaussian nature of the states while enabling entanglement generation and frequency tuning.

\subsubsection{Analytical Expression for Circuit Depth}

Our analytical derivation yields a closed-form expression for the circuit depth:
\begin{equation}
	\mathcal{D}(U) = \frac{1}{2} \log\left[\frac{A_1 A_2 - A_{12}^2}{\omega_R^2}\right] + \left|\frac{A_{12}}{A_1}\right|.
	\label{eq:depth}
\end{equation}
This expression reveals two distinct physical contributions to the total quantum complexity:
The logarithmic term captures the global scaling transformation between the reference and target states.
The linear term \( |A_{12}/A_1| \) quantifies the computational overhead arising from mode entanglement.

Notably, the off-diagonal covariance matrix element \(A_{12}\), which encodes the quantum interference between the two oscillator modes, emerges as a crucial determinant of circuit complexity. Its magnitude directly increases the depth of the required quantum circuit.
The behavior of the circuit depth varies significantly across different physical regimes, reflecting fundamental transitions in the system's quantum dynamics.
\begin{itemize}
	\item Weak Coupling Regime:	
	In the weak coupling regime (\(g \ll \omega_1, \omega_2, \omega_c\)), the mixing angle is small, suppressing the interference term \(A_{12}\). The circuit depth simplifies to:
	\begin{equation}
		\mathcal{D}(U) \approx \frac{1}{2} \log\left[\frac{A_1 A_2}{\omega_R^2}\right],
	\end{equation}
	indicating predominantly adiabatic dynamics with minimal complexity overhead. This regime is optimal for quantum operations requiring low computational resources and high fidelity.
	
	\item Strong Coupling Regime:	
	Near maximal mixing (\(\theta \approx \pi/4\)), the interference term is amplified to \(A_{12} \approx \sqrt{g/2}\), leading to:
	\begin{equation}
		\mathcal{D}(U) \approx \frac{1}{2} \log\left[\frac{A_1 A_2 - g/2}{\omega_R^2}\right] + \frac{\sqrt{g/2}}{A_1}.
	\end{equation}
	The dominant additive term signals a breakdown of adiabaticity and a substantial increase in circuit complexity, as visually demonstrated in the right panel of Figure~\ref{Fig.1}.
	
	\item External Field Dominated Regime:	
	When the magnetic field dominates the dynamics (\(\omega_c \gg \omega_1, \omega_2, g\)), the diagonal terms of the covariance matrix overwhelm the off-diagonal couplings. This suppresses mode entanglement, reducing the circuit depth to a simple logarithmic scaling:
	\begin{equation}
		\mathcal{D}(U) \approx \log\left[\frac{\omega_c}{\omega_R}\right].
	\end{equation}
	This universal logarithmic scaling is consistent with behavior observed in other quantum systems \cite{Jefferson:2017sdb}, thereby validating our analytical approach.
\end{itemize}

\subsubsection{Experimental Implications}

The system exhibits two distinct dynamical regimes: adiabatic evolution, characterized by smooth, ground-state-preserving dynamics with minimal circuit complexity (logarithmic scaling), and nonadiabatic evolution, marked by rapid state transitions where quantum interference and mode entanglement substantially increase circuit depth (linear scaling). The crossover from the adiabatic to the nonadiabatic regime is controlled by the coupling strength \(g\), external field \(\omega_c\), and detuning \(\Delta = \omega_1^2 - \omega_2^2\).

This delineation provides concrete experimental guidance: weak coupling with strong external fields enables high-fidelity, low-complexity control, ideal for robust quantum operations. In contrast, strong coupling necessitates advanced techniques to manage the heightened complexity and decoherence risks. Crucially, our results reveal a fundamental trade-off: optimizing for quantum correlations (e.g., mutual information) often incurs a direct cost in operational fidelity and complexity, a pivotal consideration for designing efficient quantum control protocols.

\begin{figure}[]
		\includegraphics[width=8cm]{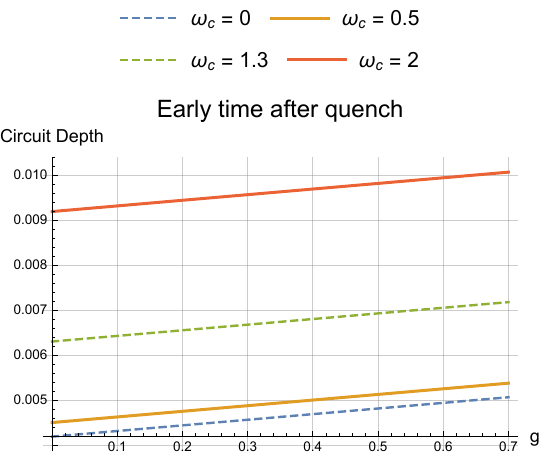} 	\includegraphics[width=8	cm]{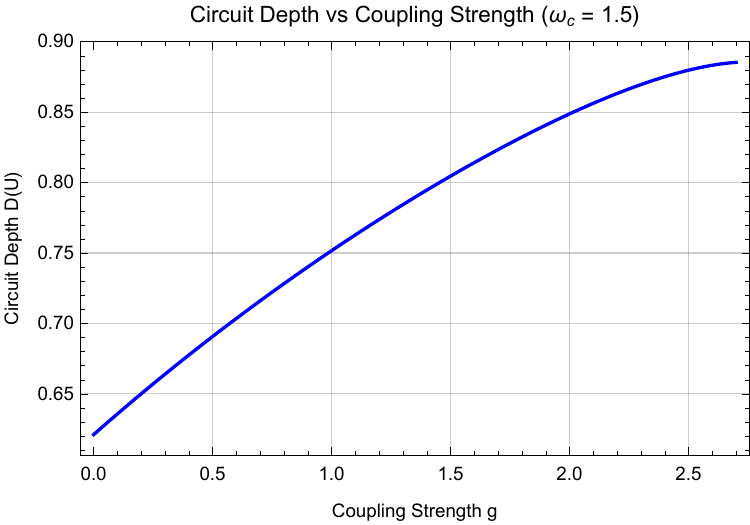}
			\caption{Left panel: Schematic diagram of the circuit depth as a function of the coupling strength $g$ at early times following the quench, where we fix $\omega_R = 1.0$, $\omega_1 = 2.0$, and $\omega_2 = 2.01$. Right panel:  In the steady state approximation circuit depth versus coupling strength $g$ is shown for $\omega_1 = 1.0$, $\omega_2 = 1.2$, $\omega_c = 1.5$, and $\omega_R = 1.0$. The curve exhibits an initial growth for small $g $. 
		 	}
		\label{Fig.1}
	\end{figure}

Figure \ref{Fig.1} comprehensively characterizes the behavior of circuit depth, across different physical regimes. The left panel demonstrates that circuit depth increases systematically with both the coupling strength \(g\) and the external magnetic field \(\omega_c\), establishing their synergistic role in driving computational overhead. The right panel reveals a fundamental dynamical transition: an initial plateau at weak coupling, where negligible interference (\(A_{12} \approx 0\)) signifies adiabatic, low-complexity dynamics, gives way to a growth phase and eventual saturation as \(g\) increases. This saturation at the theoretical limit of \(\frac{1}{2} \log(g^2/4\omega_R^2) + \frac{1}{2}\) signals maximal mode mixing (\(\theta \to \pi/4\)) and the onset of a non-adiabatic, high-complexity regime, consistent with universal complexity scaling in quantum field theories \cite{Jefferson:2017sdb}.

\begin{figure}[htbp]
		
		\includegraphics[width=0.48\textwidth]{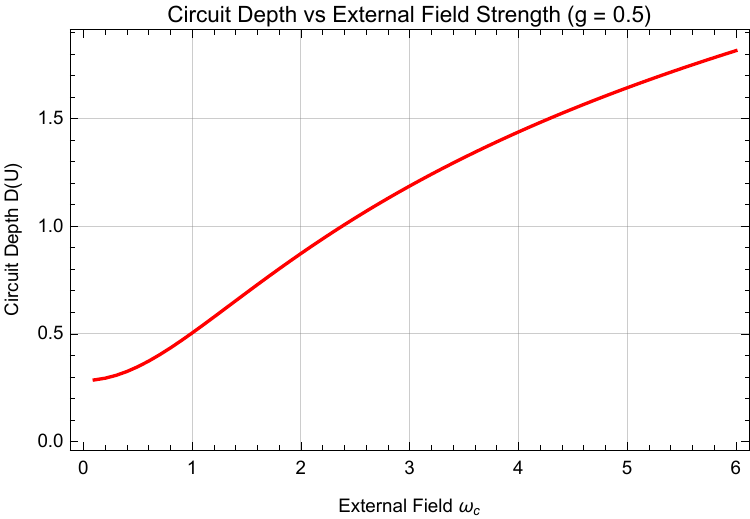}
		\includegraphics[width=0.48\textwidth]{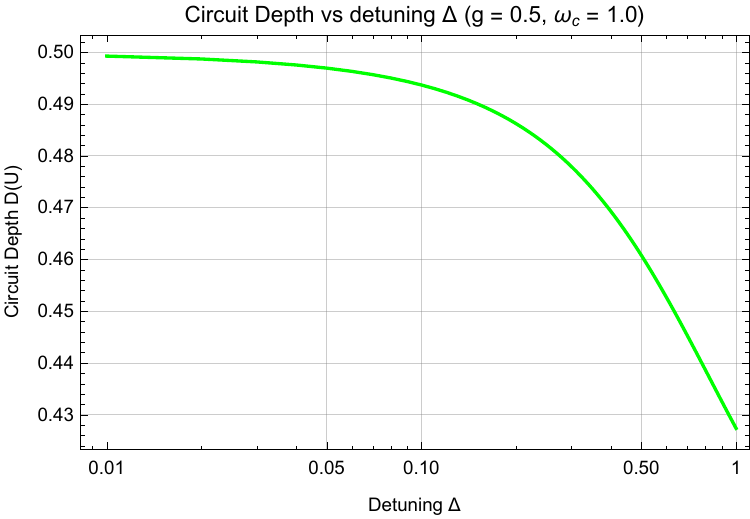}
		\caption{Left panel: Circuit depth versus external field $\omega_c$ for fixed $g = 0.5$, showing the suppression of complexity at large $\omega_c$. The curve follows the predicted $\log(\omega_c/\omega_R)$ scaling when $\omega_c$ exceeds the other energy scales. 
			 Right panel: Circuit depth versus detuning $\Delta = \omega_1^2 - \omega_2^2$ on log-log scales, 
			  for $g = 0.5$, $\omega_c = 1.0$. 
			  		}
		\label{fig:degeneracy}
	\end{figure}

This crossover, tunable via \(g\), \(\omega_c\), and the detuning \(\Delta \), provides critical experimental guidance. Furthermore, Figure \ref{fig:degeneracy} elucidates the role of external control parameters. Its left panel confirms the suppression of complexity in field-dominated regimes \(\omega_c \gg \omega_1, \omega_2, g\), where the asymptotic behavior follows the predicted \(\log(\omega_c/\omega_R)\) scaling. The right panel highlights the profound impact of detuning, showing a sharp divergence in circuit depth as \(\Delta \to 0\). This critical slowing down near degeneracy points, analogous to a phase transition, underscores the heightened sensitivity and control challenge in nearly resonant systems.

Collectively, these results map the landscape of quantum complexity, offering some protocols for optimizing control strategies in quantum technologies.




\section{Quantum Control and Transport Fidelity in a One-Body Model}
\label{sec:one_body}

Building on our analysis of quantum correlations and computational complexity in coupled systems, we now investigate quantum control in a fundamental setting: the transport of a single ion in a moving harmonic trap. This model serves as a critical testbed for understanding the fidelity-complexity trade-off in quantum operations, connecting the abstract measures from previous section.

The system is governed by the time-dependent Hamiltonian:
\begin{equation}
	\hat{H}(t) = \frac{\hat{p}^2}{2m} + \frac{1}{2} m \omega^2 \left[\hat{x}-d(t)\right]^2,
	\label{eq:Hamiltonian}
\end{equation}
where \(d(t)\) controls the trajectory of the trap minimum. To analyze the resulting dynamics, we employ a co-moving frame via the displacement operator \(\hat{D}(\alpha(t)) = \exp[\alpha(t) \hat{a}^\dagger-\alpha(t)^* \hat{a}]\). Assuming an initial ground state \(|\Psi(0)\rangle = |0\rangle\), the system evolves into a coherent state \(|\Psi(t)\rangle = |\alpha(t)\rangle\), with amplitude:
\begin{equation}
	\alpha(t) = \sqrt{\frac{m \omega}{2 \hbar}} \left(d(t) - e^{-i \omega t} \int^t_0 \dot{d}(t_1) e^{i \omega t_1} dt_1 \right).
	\label{al}
\end{equation}

We contrast two transport protocols to elucidate the interplay between control speed and quantum resource overhead:
\begin{itemize}
	\item Sudden Displacement: \(d_1(t) = d_0 \Theta(t)\), an instantaneous quench.
	\item Smooth Displacement: \(d_2(t) = L \sin^2(\pi t / 2T)\), an adiabatically-inspired profile \cite{r2}.
\end{itemize}

\subsection{Characterizing Transport: Fidelity, Complexity, and Nonadiabaticity}

We quantify transport performance using three complementary metrics that extend our analysis from previous section. Fidelity, measuring state overlap, simplifies for coherent states to:
\begin{equation}
	F(\alpha(t)) = \exp\left( -|\alpha(t) - \alpha(0)|^2 \right).
\end{equation}

To quantify computational resource costs, we require a complexity measure sensitive to displacement operations. While the circuit depth \(\mathcal{D}(U)\) from Section 3 effectively characterizes Gaussian state preparation, it yields zero for coherent states due to their identical covariance matrix with the vacuum. We therefore employ the Thermofield Double (TFD) formalism \cite{Guo:2020dsi, Chapman:2018hou, Doroudiani:2019llj}, which provides a geometric complexity measure sensitive to displacement:
\begin{equation}
	\mathcal{C}(\alpha(t)) = \vartheta \, \mathrm{csch} \left( \frac{\vartheta}{2} \right) \sqrt{ \left( |\alpha(t)|^2 + 2 \right) \cosh \vartheta - 2 },
\end{equation}
where $\vartheta = \tanh^{-1} (e^{-\beta \omega / 2})$ is the squeezing parameter encodes temperature dependence. This measure captures the operational cost associated with both displacement and finite-temperature effects, enabling direct comparison of protocol efficiency.

Finally, the nonadiabaticity parameter provides time-resolved excitation monitoring:
\begin{equation}
	\mathcal{Q}(t) = \frac{\langle \hat{H}(t) \rangle - E_0}{E_0} = 2 |\alpha(t)|^2,
\end{equation}
quantifying instantaneous deviation from adiabatic following and linking directly to the adiabatic/nonadiabatic regimes defined in Section 3.

\subsection{Protocol Performance and the Fidelity-Complexity Trade-off}

\begin{figure}[tbp]
	\includegraphics[width=0.47
	\textwidth]{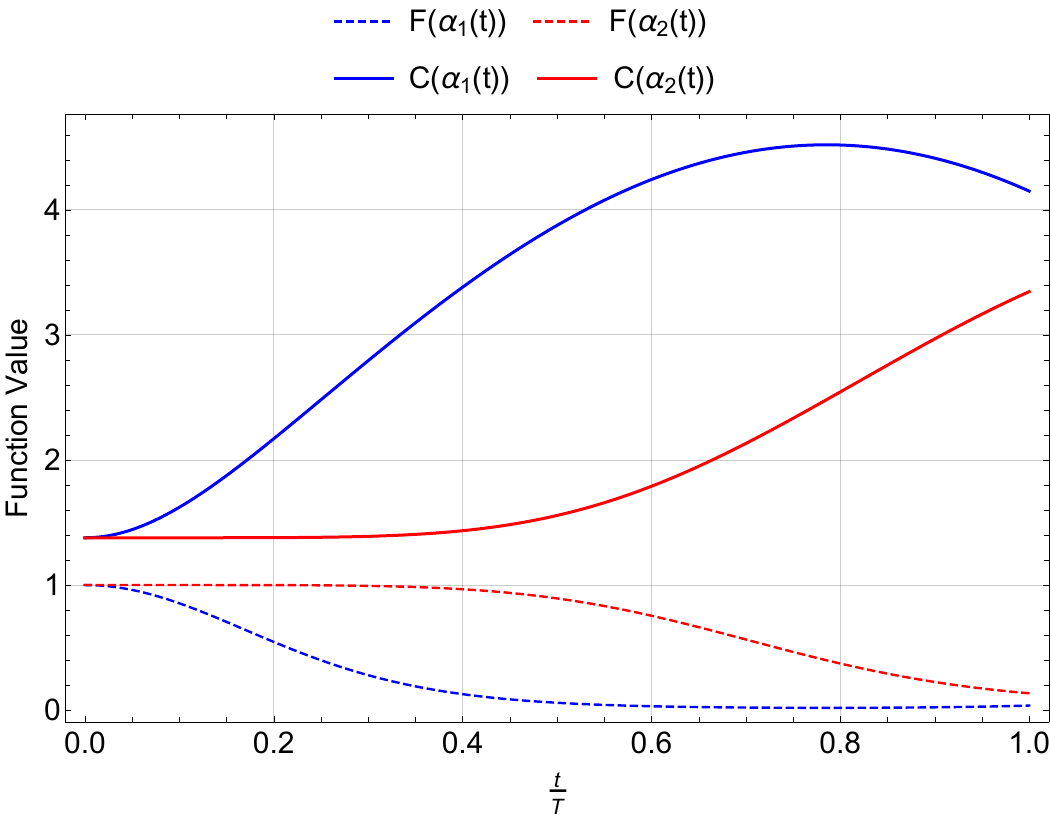} 	\includegraphics[width=0.47 	\textwidth]{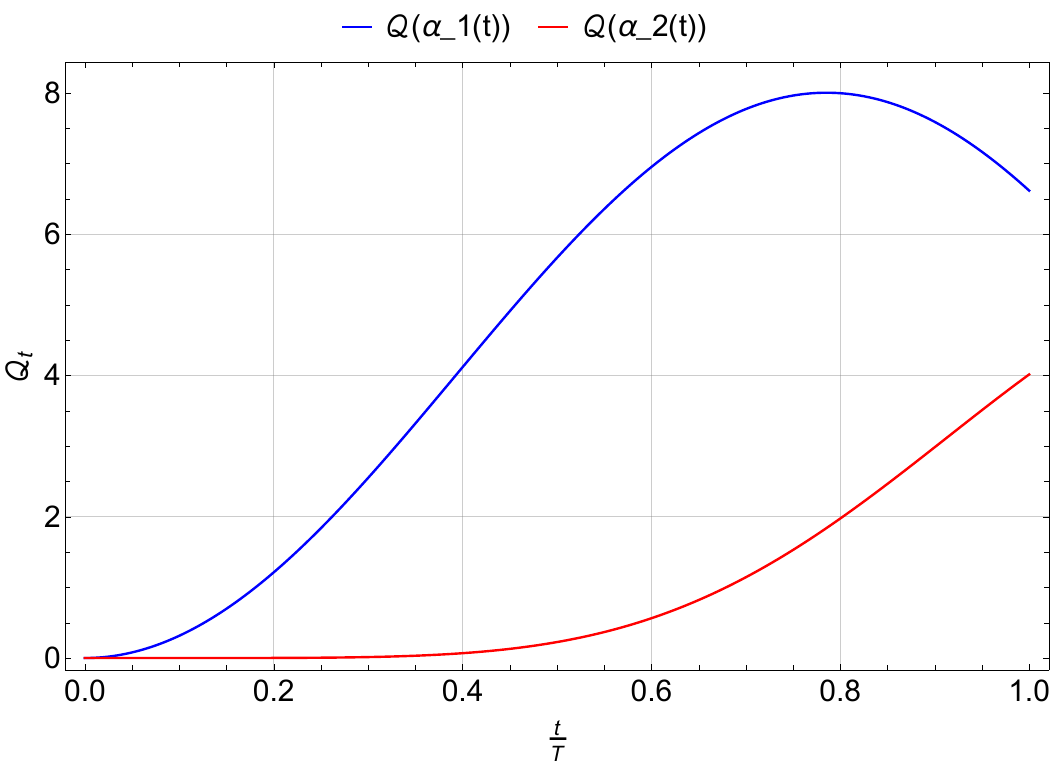}
	\caption{Left: Schematic diagram of complexity and fidelity as a function of time for two corresponding  amplitudes for two kinds of displacements.  We have set $m=d_0=L=1$, $T=2$, $\beta=1$ and $\omega=2$. Right:  Nonadiabaticity parameter $\mathcal{Q}(t)$ for two transport protocols: sudden (blue) and smooth sinusoidal (red). The sudden displacement leads to sustained excitation, while the smooth protocol suppresses transient energy buildup, indicating more adiabatic and efficient transport.		
	}	\label{fig1}
\end{figure}

\begin{figure}[tbp]
	\includegraphics[width=0.5\textwidth]{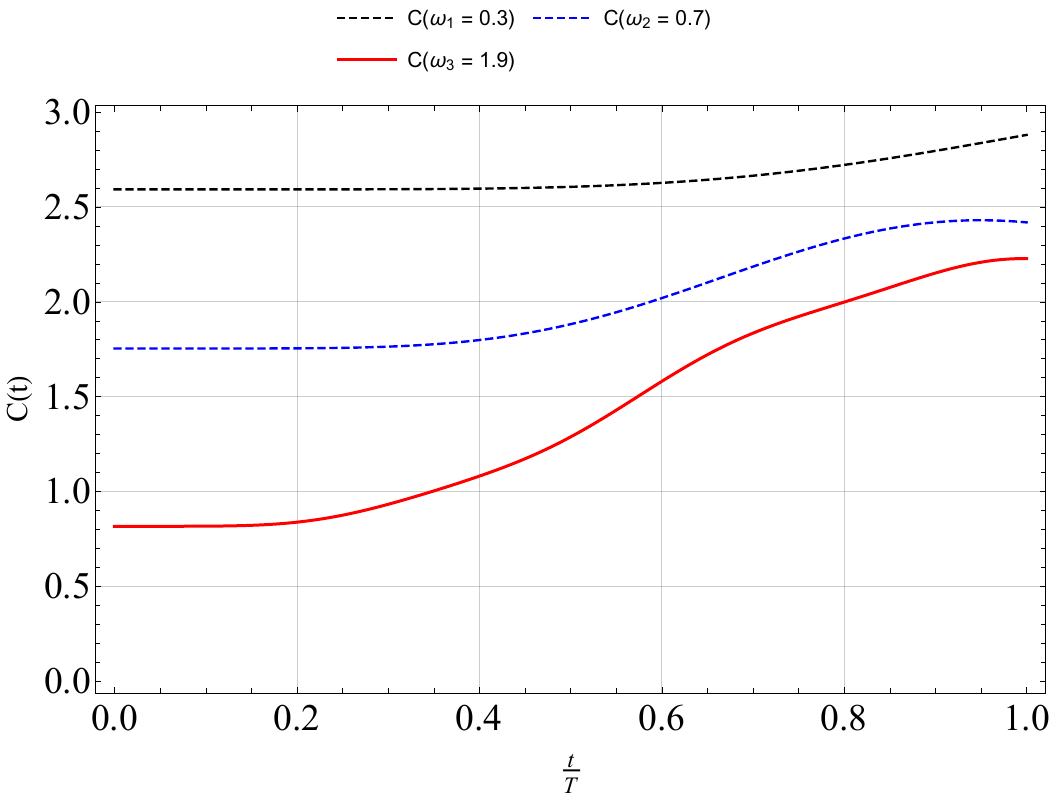}\includegraphics[width=0.5\textwidth]{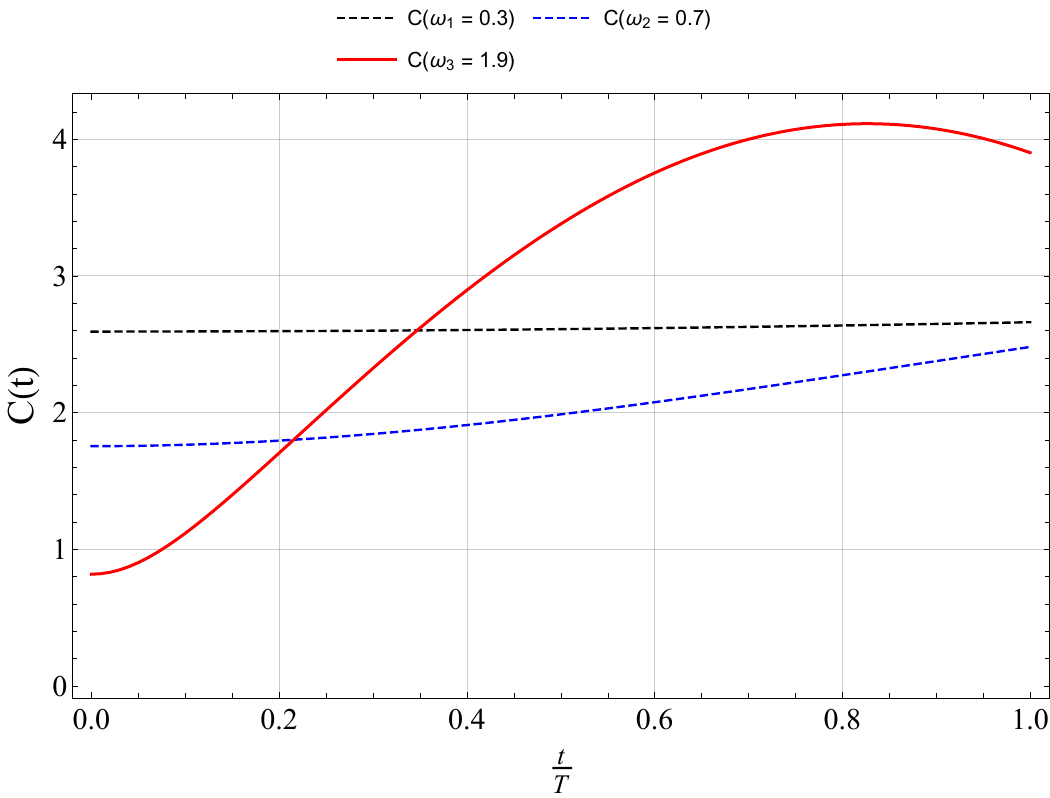}		
	\caption{Schematic diagram of complexity as a function of time for different values of $\omega$: Left: for the amplitude of $\alpha_2(t)$   Right: for $\alpha_1(t)$. We have set $m=d_0=L=1$ and $T=2$, $\beta=1$ }	\label{fig0}
\end{figure}

Analysis of these metrics (Figures \ref{fig0} and \ref{fig1}) reveals a fundamental control principle. The sudden quench protocol generates significant, persistent excitations, reflected in sustained high values of \(\mathcal{Q}(t)\) and correspondingly large complexity costs—the single-body analogue of the nonadiabatic, high-complexity regime in coupled oscillators. In contrast, the smooth protocol suppresses excitations through small, transient \(\mathcal{Q}(t)\) peaks, achieving higher fidelity with lower complexity, mirroring the adiabatic, low-complexity regime.

This analysis crystallizes a key quantum control principle: smooth, temporally extended control fields suppress nonadiabatic excitations, thereby minimizing the quantum complexity required for high-fidelity operations. This provides concrete, actionable guidance for optimizing quantum control sequences in experimental platforms, demonstrating that abstract complexity measures have direct, measurable consequences for operational performance.

The insights obtained from both models build upon prior works (see for example \cite{Khorasani:2021zus, Khorasani:2023usq, 234, REZA:tanhayi, Margolus:1997ih, Deffner:2017cxz}), demonstrating the influence of external electric and magnetic fields on circuit complexity and coherence in quantum systems. Moreover, our framework offers potential applications in quantum error correction, where the ability to minimize circuit depth without compromising fidelity is essential for building scalable and fault-tolerant quantum processors.

This analysis crystallizes a fundamental quantum control principle extracted from our metrics: smooth, temporally extended control fields suppress nonadiabatic excitations, thereby minimizing the quantum complexity required for high-fidelity operations. The stark contrast between protocols demonstrates that the abstract adiabatic and non-adiabatic regimes, identified in the coupled-oscillator model, have quantifiable consequences for operational performance. This provides guidance for optimizing quantum control sequences, specifically, that minimizing the time-derivatives of control fields is a primary method for reducing computational overhead. By bridging abstract complexity measures and practical fidelity, our framework offers a direct pathway to improving resource-efficiency in critical applications like quantum error correction, where minimizing circuit depth without compromising fidelity is essential for scalable, fault-tolerant quantum processors.

\section{Conclusion }
This work establishes a unified framework for quantifying the interplay between quantum correlations, coherence, and computational complexity through two complementary models. Our central finding reveals a fundamental dichotomy: measures of total correlation (mutual information) and dynamical coherence (synchronization) exhibit distinct and often opposing responses to system parameters, challenging the notion of a universal concordance between them.

In the coupled-oscillator model, we have demonstrated that mutual information and synchronization respond differently to key parameters: stronger coupling enhances shared information but suppresses coherent phase alignment, and external magnetic fields further amplify this split. This shows that greater quantum correlation, as measured by mutual information, does not ensure stronger synchronization, highlighting a clear distinction between informational and dynamical measures of correlation. Furthermore, we quantified the computational cost of these correlations through circuit complexity, which undergoes a sharp transition from adiabatic (low-depth) to non-adiabatic (high-depth) regimes. The saturation of complexity at the strong-coupling limit and its logarithmic suppression by external fields provide clear, universal scaling laws for resource estimation.

The single-ion transport model translated these abstract measures into a concrete fidelity-complexity trade-off. We showed that smoothly modulated control protocols suppress nonadiabatic excitations, simultaneously maximizing fidelity and minimizing complexity, while sudden quench protocols incur high computational overhead. This provides an actionable principle for quantum control: temporal smoothness is a resource for efficiency.

These insights have immediate implications for quantum technologies. The delineation of low-complexity regimes offers blueprints for optimizing quantum circuits in platforms like trapped ions and superconducting qubits. The observed fidelity-complexity trade-off is directly relevant to quantum error correction, where minimizing circuit depth is essential for fault tolerance.

Future work should extend this framework to non-Gaussian states and multi-particle entangled systems, where richer synchronization phenomena and complexity scaling are expected. Experimental validation of the predicted complexity saturation and control principles would firmly bridge our theoretical results with practical quantum engineering. Ultimately, by mapping the intricate landscape between information, coherence, and complexity, this work provides a refined toolkit for steering quantum dynamics toward both correlated and computationally efficient states.

\subsection*{Acknowledgments}

We would like to express our gratitude to Mohsen Alishahiha, Mohammad Vasli, Sadaf Ebadi, and M. Hossein Khoshnevis for their insightful discussions and valuable feedback on various aspects of this work. We also appreciate the constructive comments from the anonymous reviewers, which helped clarify and improve the manuscript. Additionally, R. Pirmoradian would like to thank M. Reza Lahooti for his ongoing support. This work is based upon research founded by Iran National Science Foundation (INSF) under project No 4026389.

	\end{document}